\documentclass[12pt]{article}

\usepackage{amsmath}
\usepackage{times}
\usepackage{subfigure}
\usepackage{graphicx}
\usepackage{comment}
\usepackage{lineno}
\usepackage{fancyhdr}
\usepackage[font=small,labelsep=none]{caption}
\newenvironment{myabstract}{%
\begin{quote} }
{\end{quote}}

\title{Inverse Design of Fluid Flow Structure with Turing Pattern}

\author
{Ercan M. Dede,$^{1\ast}$ Yuqing Zhou,$^{1}$ Tsuyoshi Nomura$^{1,2}$ \\
\\
\normalsize{$^{1}$Electronics Research Department, Toyota Research Institute of North America,}\\
\normalsize{1555 Woodridge Ave., Ann Arbor, MI 48105, USA}\\
\normalsize{$^{2}$Toyota Central R\&D Labs, Inc., 41-1 Yokomichi, Nagakute 480-1192, Japan}\\
\\
\normalsize{$^\ast$Corresponding author; E-mail: eric.dede@toyota.com}
}
\date{}

\begin{document} 
\baselineskip24pt
\maketitle 

\begin{myabstract}
Microchannel reactors are critical in biological plus energy-related applications and require meticulous design of hundreds-to-thousands of fluid flow channels. Such systems commonly comprise intricate space-filling microstructures to control the fluid flow distribution for the reaction process. Traditional flow channel design schemes are intuition-based or utilize analytical rule-based optimization strategies that are oversimplified for large-scale domains of arbitrary geometry. Here, a gradient-based optimization method is proposed, where effective porous media and fluid velocity vector design information is exploited and linked to explicit microchannel parameterizations. Reaction-diffusion equations are then utilized to generate space-filling Turing pattern microchannel flow structures from the porous media field. With this computationally efficient and broadly applicable technique, precise control of fluid flow distribution is demonstrated across large numbers (on the order of hundreds) of microchannels.
\end{myabstract}

\clearpage

\section{Introduction}
\label{intro}
Microchannel flow structures are found in a range of important industrial applications involving water purification~\cite{Wang14}, pharmaceuticals~\cite{gutmann15}, electronics~\cite{CHEN20022643}, and green chemistry~\cite{LEROU2010380}. Such systems require careful handling of fluid species to control reaction processes. Particularly, the flow distribution of fluids is critical, and intricate space-filling channel structures are employed for flow control through single- or multi-layered (e.g. 100-10,000) microchannels.

Design optimization schemes for uniform fluid delivery to arrays of microchannels include approximation techniques, heuristic strategies, bifurcation methods, and gradient-based algorithms; see~\cite{rebrov2011single}. One approximation technique,~\cite{commenge02}, treats flow friction via a resistive network of ducts to understand pressure drop across a flow distribution chamber (or manifold). Using analytical expressions, the influence of the manifold geometry is understood, and flow uniformity is optimized for simplified (e.g. trapezoidal) geometries. A heuristic technique,~\cite{LUO2015542}, involves perforated baffles upstream of microchannels to homogenize fluid distribution, although added pressure drop is a concern. The use of constructal theory,~\cite{Bejan97}, is popular with some,~\cite{Senn04}, using tree-like flow distributors for fuel cells and others,~\cite{TONDEUR20041799}, applying fractal manifolds to hundreds of channels in an adsorbent monolith. An associated challenge may be the spatial requirements for series configured branching.

Borrvall et al.~\cite{Borvall03} pioneered gradient-based topology optimization for Stokes flow, and this was extended to higher Reynolds numbers~\cite{Gersborg-Hansen2005,olesen06}. The method was employed,~\cite{dede14}, to optimize manifolds with less than 10 fluid outlets in an electronics heat sink. Recently, researchers,~\cite{Liu11,8419642}, incorporated mass flow rate constraints for a slightly greater number of outlets, e.g. $\sim$20. We leverage these formulations, where Navier-Stokes laminar incompressible fluid flow in an idealized porous medium is assumed with a friction force proportional to the fluid velocity, viz. Darcy's law~\cite{olesen06,kaviany95},
\begin{equation}
    \nabla \cdot \mathbf{u} = 0, \quad \mbox{and}
    \label{eqn:incon}
\end{equation}
\begin{equation}
    \rho \left( \mathbf{u} \cdot \nabla\mathbf{u} \right) = -\nabla P + \nabla \cdot \left\{ \eta \left[ \nabla \mathbf{u} + \left( \nabla \mathbf{u} \right)^{\mbox{\small T}} \right] \right\} - \eta \alpha\left(\gamma \right) \mathbf{u}.
    \label{eqn:ns}
\end{equation}

\noindent Fluid incompressibility is governed by~(\ref{eqn:incon}), and flow in the idealized porous medium is defined by~(\ref{eqn:ns}). The fluid pressure and velocity vector state variables are given by $P$ and $\mathbf{u}$, respectively. The fluid dynamic viscosity is $\eta$. The effective inverse permeability is a function of a design variable, $\gamma$, and in standard formulations,~\cite{olesen06}, is interpolated using the convex function $\alpha=\alpha_s$,
\begin{equation}
    \alpha_s \left(\gamma \right) = \alpha_{\mbox{\small min}} + \left( \alpha_{\mbox{\small max}} - \alpha_{\mbox{\small min}} \right) \frac{q(1-\gamma)}{q+\gamma}.
    \label{eq:std_interp}
\end{equation}
A low permeability quasi-solid state exists for $\gamma \rightarrow 0$ by $\alpha_{\mbox{\small max}} = 1 / l^2 Da$ , where $Da$ is the Darcy number, and $l$ is the characteristic length. A fluid state, where the friction force term goes to zero when $\gamma \rightarrow 1$, emerges with $\alpha_{\mbox{\small min}} = 0$. The goal in adopting (\ref{eq:std_interp}) with the convex tuning parameter, $q$, is to obtain completely fluid or solid states. However, challenges arise in selecting the appropriate Darcy number, $Da$, to eliminate gray-scale designs that permit undesirable flow seepage through quasi-solid material.

For a design space, $\Omega$, an accepted~\cite{Borvall03,olesen06,Gersborg-Hansen2005,dede14} objective function, $f_o$, is to minimize power dissipation or flow resistance, 
\begin{equation}
    f_o = \int_\Omega{\left[ \frac{1}{2} \eta \sum_{i,j}{\left(\frac{\partial u_i}{\partial x_j} + \frac{\partial u_j}{\partial x_i} \right)^2 + \eta \sum_{i}{\alpha_s \left(\gamma \right) u_i^2}} \right]} \mbox{d}\Omega.
    \label{eq:obfun}
\end{equation}

A volume constraint is typical to control the amount of fluid in the result~\cite{olesen06,Gersborg-Hansen2005,dede14}. However, space-filling structures are not obtainable since channel-to-wall spacing is not enforced. In contrast, removal of the volume constraint produces large open flow channel designs to naturally minimize flow resistance.

For uniform mass flow through discrete fluid outlets, a mass flow rate constraint is common~\cite{Liu11,8419642},
\begin{equation}
    \left( \frac{\dot{m}_k}{\dot{m}_{t,k}} - 1 \right)^2 \leq \delta^2 \quad \mathrm{for} \hspace{1mm} k=1,2,3,\ldots,n, 
    \label{eq:mfrconst}
\end{equation}
where $\dot{m}_k$ and $\dot{m}_{t,k}$ are the individual and target mass flow rate, respectively, at the $k^{th}$ outlet, and $\delta$ is an outlet flow rate error tolerance. To represent large gradients in the flow solution (e.g. near outlets), many channels require an extremely fine computational grid, per~\cite{reddy}. Thus, effective optimization strategies for fluid flow control to hundreds of microchannels in application is still a critical field of research. Complex space-filling microchannel patterns are common, and for arbitrary domains, configurations may not be expeditiously found using approximation, rule-based, or modern gradient-based techniques. 

Complex patterns also exist in nature at multiple scales (e.g. mammalian markings, fish skin, seashells, etc.), and reaction-diffusion equations replicate irregular spatio -- temporal Turing patterns~\cite{Pearson189,gs85,GARIKIPATI2017192,Kondo1616}. Here, we combine gradient-based porous media optimization in an arbitrary fluid flow domain with a reaction-diffusion model for computationally efficient development of intricate space-filling microchannel architectures. Explicit modeling of channels is abandoned, and channel synthesis is realized through a Turing pattern generation algorithm.

\section{Methodology}
\begin{figure}[t!]
    \centering
    \includegraphics[]{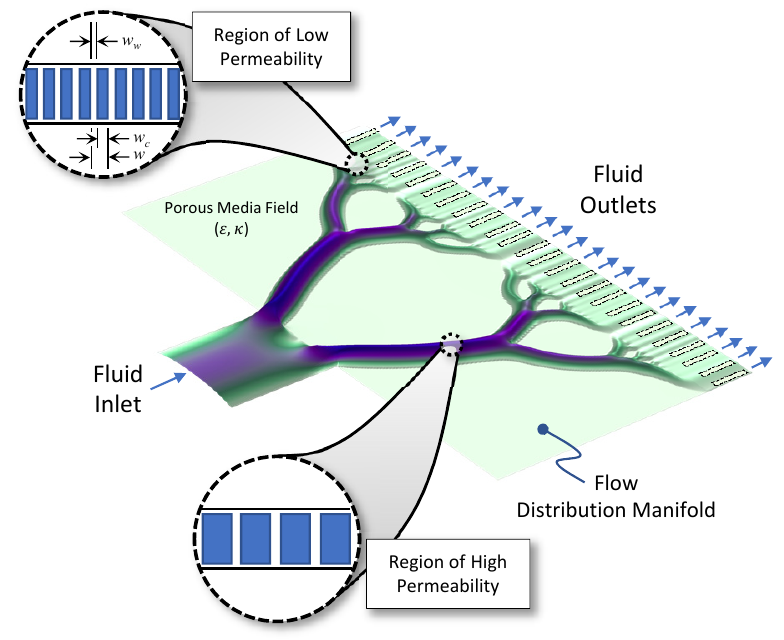}
    \caption{\quad 2-D flow distribution manifold concept; the porous media has porosity, $\epsilon$, and permeability, $\kappa$, that vary spatially and are parameterized by the microstructure wall width, $w_w$, channel width, $w_c$, and channel plus wall spacing, $w$.}
    \label{fig:concept}
\end{figure}

The approach comprises two steps. First, to scale up to manifolds with hundreds of outlets, the problem is re-framed to design a homogenized porous fluid flow structure, where all material states are physically feasible. Following~\cite{Kim99}, the porous media is parameterized in two-dimensions (2-D) based on a spatially varying local microchannel structure; see Fig.~\ref{fig:concept}. The porosity, $\epsilon$, and permeability, $\kappa$, are,

\begin{equation}
    \epsilon = \frac{w_c}{w}, \quad \kappa = \frac{\epsilon w_c^2}{12}, \quad \mathrm{with} \quad w=w_c+w_w,
    \label{eqn:kim}
\end{equation}
where $w_c$ and $w_w$ are the channel and wall widths, respectively. 

A linear interpolation function for the channel width relates any porous media state to a microstructure,
\begin{equation}
    w_c \left(\gamma \right) = w_{c\mathrm{min}} + \left( w_{c\mathrm{max}} - w_{c\mathrm{min}} \right)\gamma,
    \label{eqn:linint}
\end{equation}
where $w_{c\mathrm{min}}$ and $w_{c\mathrm{max}}$ are minimum and maximum microchannel widths, respectively.

Combining~(\ref{eqn:kim}) and (\ref{eqn:linint}), and assuming $w_w$ is constant, a new inverse permeability expression, $\alpha \rightarrow \alpha_n$, is,
\begin{equation}
    \alpha_n \left(\gamma \right) = \frac{1}{\kappa(\gamma)} =  12\left[\frac{1}{w_c(\gamma)^2}+\frac{w_w}{w_c(\gamma)^3}\right].
\end{equation}
Thus, 0-1 (fluid-solid) designs are not required, flow seepage is not a concern, a volume constraint is unnecessary, and hundreds of discrete fluid flow outlets are exchanged for an aggregated fluid outlet with specified flow profile.

The second step of the algorithm to find the Turing pattern is an expansion of the anisotropic thermal-composite design approach explained in~\cite{Petrovic18}. The Turing reaction-diffusion system is a mathematical model of the morphogenesis of the embryo proposed by A.M. Turing~\cite{turing1952chemical}, involving two interacting hypothetical chemical substances U and V, which diffuse in the space around and enhance or suppress the reproduction of themselves~\cite{Kondo1616}. 
The dimensionless equations for this process are,
\begin{eqnarray}
    \frac{\partial U}{\partial t} &= D_u \nabla^2 U +R_u(U, V),\\
    \frac{\partial V}{\partial t} &= D_v \nabla^2 V + R_v(U, V),
\end{eqnarray}
where $R_u(U, V)$ and $R_v(U, V)$ are interactive reaction terms. 
In this study, the reaction terms are augmented following~\cite{Kondo1616,Petrovic18};
\begin{eqnarray}
    R_u(U,V)&=(a_uU+b_uV+c_u)-d_uU=F(U,V)-d_uU,\\    
    R_v(U,V)&=(a_vU+b_vV+c_v)-d_vV=G(U,V)-d_vV,
   \end{eqnarray}
    where
\begin{eqnarray}
    0\leq F(U,V)=a_uU+b_uV+c_u\leq F_{\mathrm{max}},\\
    0\leq G(U,V)=a_vU+b_vV+c_v\leq G_{\mathrm{max}}.
\end{eqnarray}

For 2-D flow systems, we extend the diffusion coefficients as 2 $\times$ 2 tensors, $\mathbf{D}_u$ and $\mathbf{D}_v$, with anisotropic diffusion terms derived from local permeability field values and perturbed over time between weakly and strongly anisotropic states. By aligning the principal axis of the diffusion tensors with the fluid velocity vector, the microstructure pitch and length are controlled with length periodically elongated along the fluid flow direction.

For example, $\mathbf{D}_u$ is written using the normalized fluid flow velocity vector, $\mathbf{\bar{u}}$,
\begin{equation}
    \mathbf{D}_u(\mathbf{\bar{u}})  = (L_u - W_u)  \mathbf{\bar{u}}\otimes\mathbf{\bar{u}} + W_u \delta_{ij},
    \label{eq:dudva2}
\end{equation}
where $\otimes$ is the dyadic product operator. The coefficients in~(\ref{eq:dudva2}) are defined as $L_u=(l_u W_u)^2$ and $W_u=(w_uw)^2$ with $l_u$ set to control the magnitude of anisotropy and $w_u$ specified as a constant value that associates the channel pitch and generated microstructure pattern. By specifying channel pitch, $w$, the lateral component of $\mathbf{D}_u$, i.e. $W_u$, is proportional to $w^2$, and thus, we recover the porous media permeability distribution.

\begin{figure}[t!]
    \centering
    \subfigure[]{\includegraphics[width=30mm]{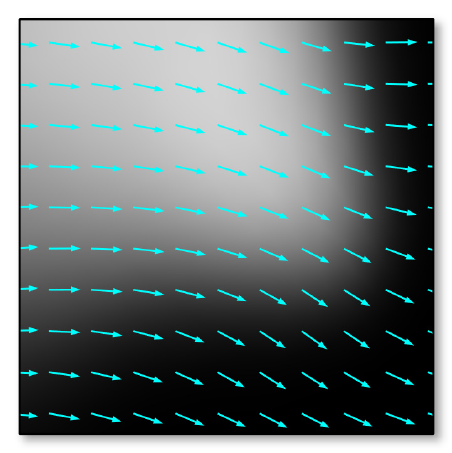}}
    \subfigure[]{\includegraphics[width=30mm]{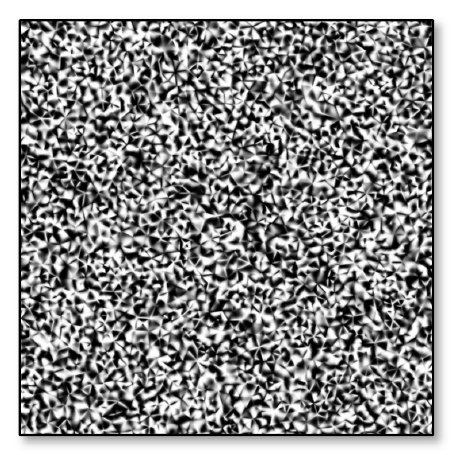}}
    \subfigure[]{\includegraphics[width=30mm]{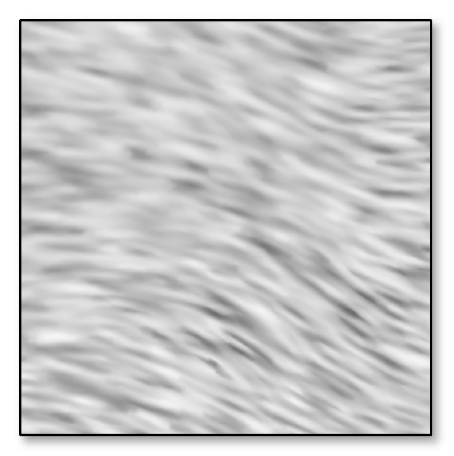}}\\
    \subfigure[]{\includegraphics[width=30mm]{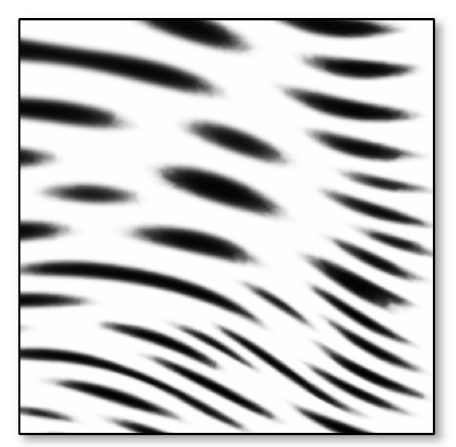}}
    \subfigure[]{\includegraphics[width=30mm]{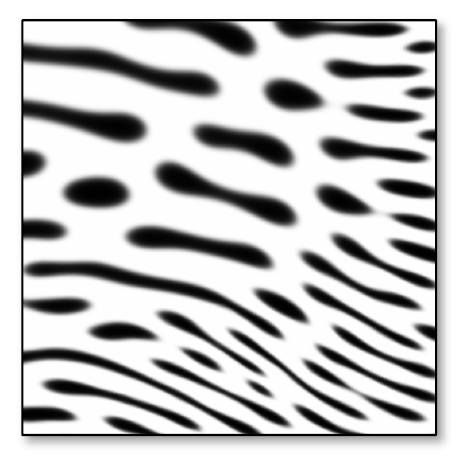}}
    \caption{\quad Turing microstructure evolution: (a) gray-scale porous medium with fluid velocity vectors (light blue), where lighter regions indicate greater permeability; (b) random, $t=0$, condition for reaction-diffusion algorithm; (c) $t=100$ s, with anisotropic diffusion coefficients; (d) $t=1000$ s, with anisotropic coefficients; (e) $t=1800$ s, with isotropic coefficients.}
    \label{fig:gs}
\end{figure}

The magnitude of the Turing pattern anisotropy repetitively varies temporally by changing $l_u$ between $\sim$1 and 10. Larger anisotropy generates a pattern strongly affected by the fluid velocity vector field, and smaller anisotropy generates a new more tightly packed pattern based on the previous final state. Figures~\ref{fig:gs}(a)-(e) show an example with an optimized fluid flow velocity field in an effective porous medium with varying permeability and the evolution of the Turing pattern microstructure. The algorithm beneficially combines homogenization of the design space for the computationally expensive flow field optimization with efficient dehomogenization of the porous media. This dehomogenization process may uniquely vary; however, to render the structure on to a physical scale, a self-organizing system based on Turing patterns is one option.

\subsection{Implementation}
The algorithm is implemented in COMSOL v.5.3a. The optimization objective function is set to minimize an equally weighted linear combination of two terms including (\ref{eq:obfun}) and variation of the fluid velocity normal to the outlet boundary. Channel and wall width values are specified to parameterize the effective inverse permeability, $\alpha_n$, of the flow space. The flow optimization uses a Method of Moving Asymptotes (MMA) optimizer that converges within $\sim$100 iterations. From the porous media field results, the anisotropic diffusion coefficient tensors for the reaction-diffusion equations are determined, and the equations are propagated through time ($\sim$1800 sec.) to generate the Turing pattern microstructure.

\section{Results}
\begin{figure}[t!]
    \centering
    \subfigure[]{\includegraphics[width=0.40\textwidth]{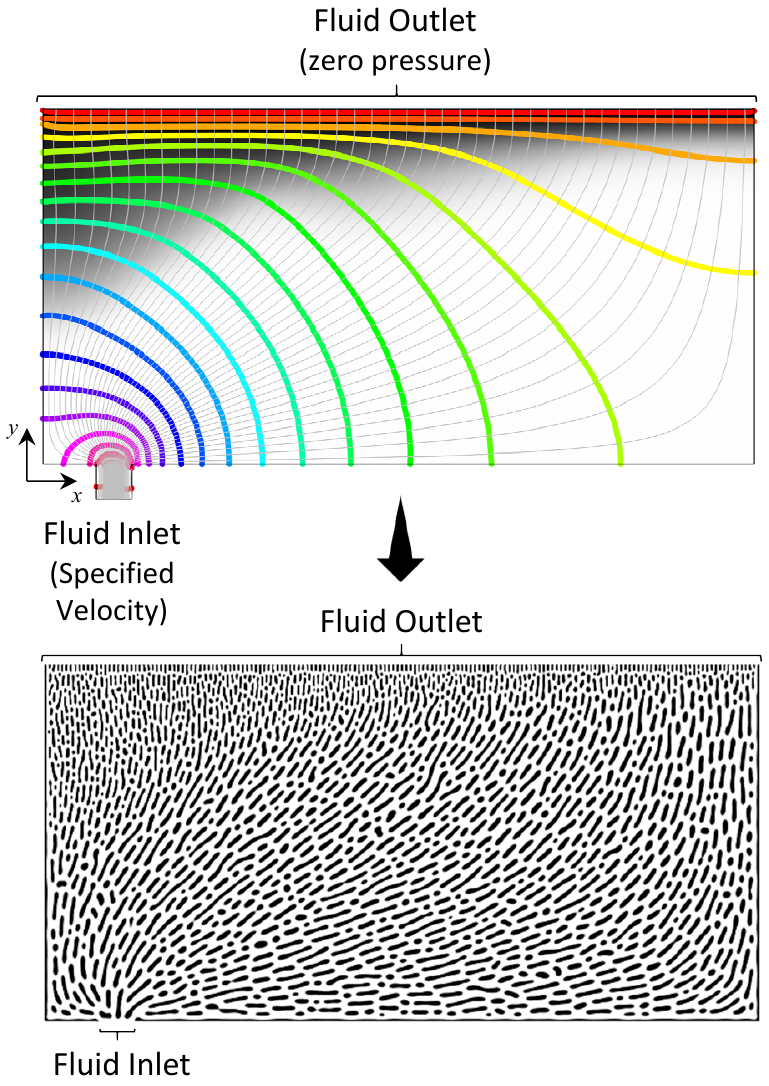}}\quad
    \subfigure[]{\includegraphics[width=0.55\textwidth]{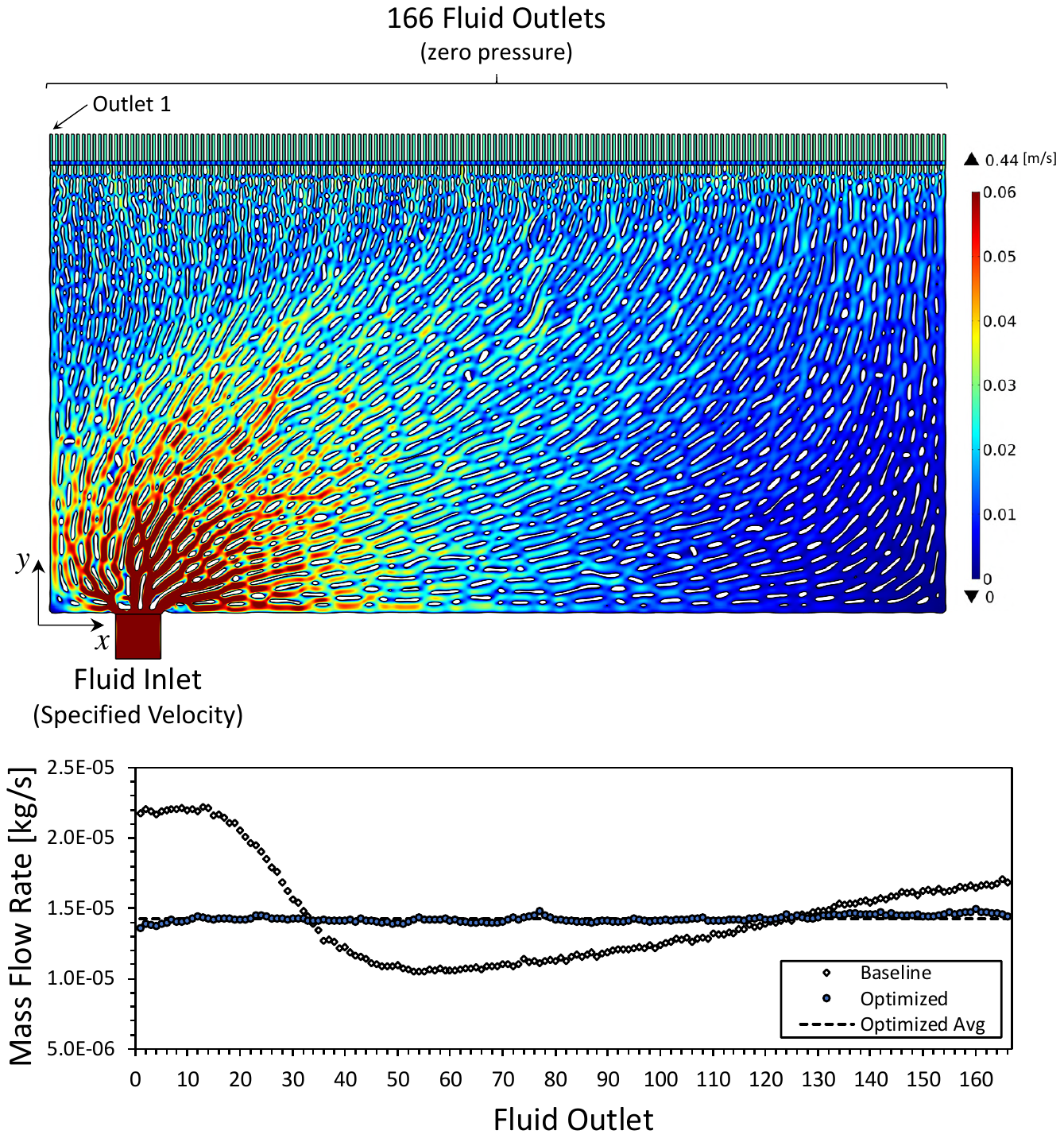}}
    \caption{\quad (a) Top - porous media optimization result with streamlines (gray) and normalized pressure contours (zero pressure at outlet) for 200 $\times$ 100 mm$^2$ design domain; light colored regions have greater permeability. Bottom - Turing pattern microstructure; light regions = fluid, black regions = solid. (b) Fluid velocity contours (top). Mass flow distribution with average outlet variation of 1.3\% for 166 fluid outlets (bottom, optimized data). Maximum variation at a single outlet is 4.6\% (optimized) versus 52.9\% for a flow domain without any microstructure features (baseline).}
    \label{fig:result1}
\end{figure}

A 200 $\times$ 100 mm$^2$ 2-D space is considered in the upper image of Fig.~\ref{fig:result1}(a) in a first numerical experiment with air flow at 20 $^\circ$C. A uniform $+y$-direction fluid velocity of 0.2 m/s is fixed over the 10 mm wide fluid inlet positioned 15 mm from the domain lower left corner. This asymmetric inlet-to-outlet configuration represents typical microchannel reactors and produces significant outlet flow maldistribution; see~\cite{commenge02}. A zero pressure outlet boundary condition (BC) is applied along the top edge of the flow space. A Dirichlet BC for the reaction-diffusion model is applied along the same top edge of the domain and on a horizontal inner boundary slightly (3 mm) below the top edge to enforce a precise channel width distribution. A second Dirichlet BC is applied on the remaining domain boundaries. The minimum and maximum channel width is $w_{c\mathrm{min}} = 0.6$ mm and $w_{c\mathrm{max}} = 1.8$ mm, respectively. The wall width between channels is uniformly fixed to $w_w = 0.6$ mm. Thus, porosity, $\epsilon$, ranges from 0.5 to 0.75 throughout the domain, while permeability, $\kappa$, ranges from 1.5E-8 m$^2$ to 2.025E-7 m$^2$. The channel width at the top outlet boundary is set to 0.6 mm, and $\sim$166 uniform width outlets are implicitly defined. 

The gray-scale fluid flow optimization result for the effective porous media is shown on top in Fig.~\ref{fig:result1}(a) including streamlines and normalized pressure contours with the Turing pattern microstructure shown below. Note that intricate patterned island wall structures are built following the above parameters. Fluid velocity contours from a flow verification analysis under the same inlet/outlet BCs are shown in Fig.~\ref{fig:result1}(b). Observe that the average variation,
\begin{equation}
    (1/n)\sum^{n}_{k=1} \mid ( \dot{m}_k - \dot{m}_{avg})/\dot{m}_{avg} \mid,
\end{equation}
in the mass flow at each outlet, $\dot{m}_k$, is very low and within 1.3\% of the mass flow average, $\dot{m}_{avg}$, across the $n=166$ outlets; the maximum variation, max$[ \mid ( \dot{m}_{k} - \dot{m}_{avg})/\dot{m}_{avg} \mid ]$, at a single outlet is 4.6\%. For comparison, the maximum variation in $\dot{m}_k$ for the flow domain without any microstructures is 52.9\%; see Fig.~\ref{fig:result1}(b).

\begin{figure}[t!]
    \centering
    \includegraphics[width=0.55\textwidth]{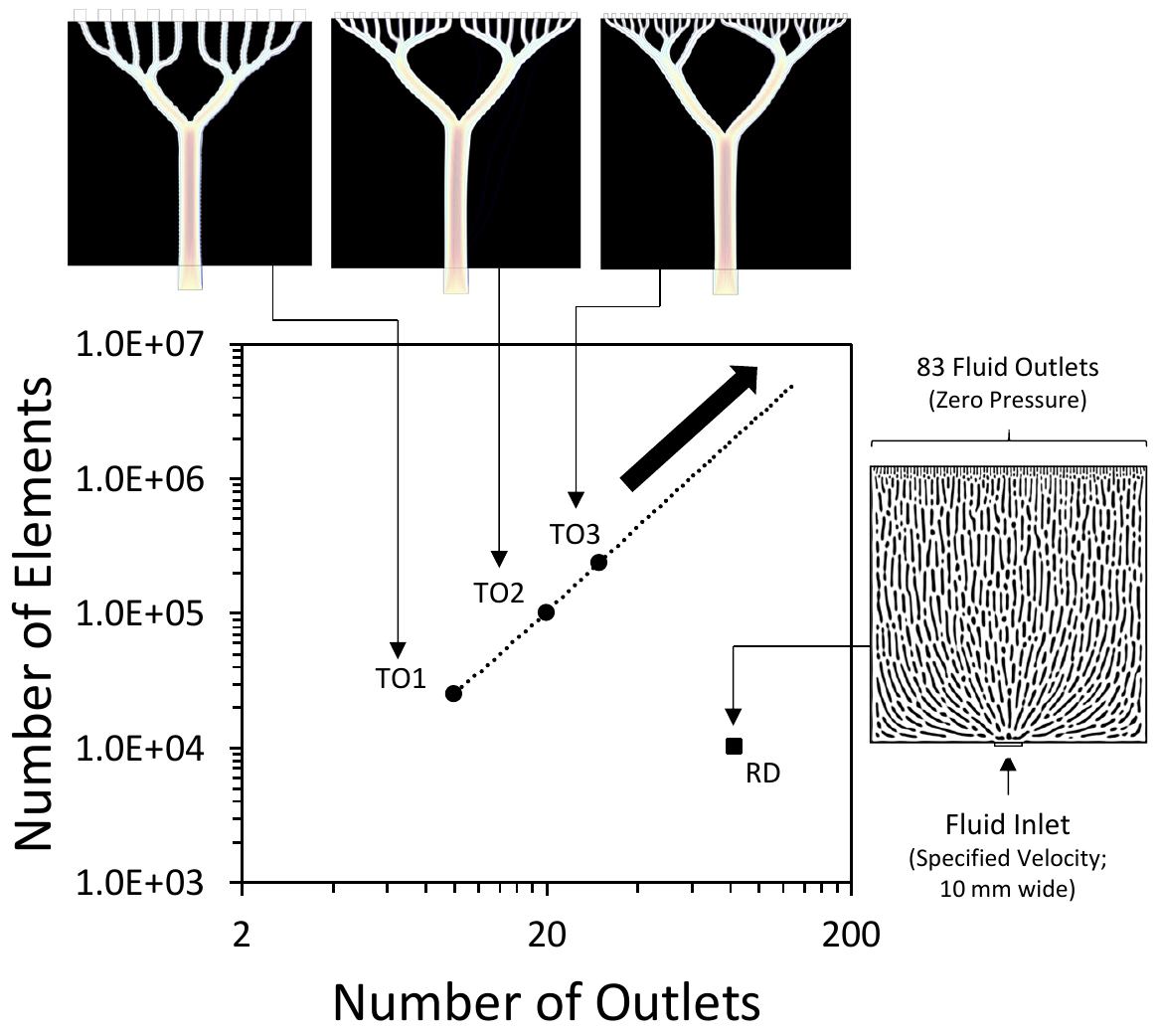}
    \caption{\quad Number of elements versus outlets for the forward solution finite element analysis of a 100 $\times$ 100 mm$^2$ manifold. Note: TO = conventional topology optimization algorithm; RD = reaction-diffusion algorithm. Light regions = fluid, black regions = solid. For TO1-TO3, the outlet mass flow rate constraint is satisfied to $<6$\%.}
    \label{fig:result2}
\end{figure}

To compare the computational effort involved in our approach with conventional flow channel topology optimization, cf.~\cite{olesen06,Liu11,8419642}, a simple 100 $\times$ 100 mm$^2$ 2-D flow space is considered in a second numerical experiment. Here, the 10 mm wide fluid inlet is positioned at the center of the lower edge of the domain. A uniform $+y$-direction fluid inlet velocity of 0.2 m/s is again assumed; the remaining BCs follow the prior example. The same microchannel and wall widths are further assumed for 83 implicitly defined outlets; the optimized Turing pattern microstructure is shown in Fig.~\ref{fig:result2}. Three flow structures from conventional topology optimization (TO1, TO2, and TO3) are also shown with 10, 20, and 30 explicitly defined fluid outlets obtained using (\ref{eq:mfrconst}). For explicitly defined outlets, the computational cost scales linearly on a log-log scale. However, the number of elements required to solve for the porous media flow field in a design of any number of implicitly defined outlets remains constant, assuming an initially sufficiently refined mesh to ensure solution accuracy. Thus, our computational procedure is efficient and additionally bounds the minimum and maximum channel widths throughout the flow domain.

\section{Conclusions}
An optimization method was proposed to design space-filling Turing patterned microstructures for fluid flow control through a porous media field. Reaction-diffusion equations form the basis of the microchannel dehomogenization post-processing technique, where the porous media field is linked to explicit representation of the rendered microstructure. Numerical experiments highlight the capability to produce precise flow control for manifolds involving hundreds of microchannel outlets. The method is effective, flexible and may be applied to large arbitrary geometries. Other porous media parameterizations are feasible opening opportunities for design in three-dimensions. Relevant uses of this methodology include the design and additive fabrication of microchannel reactors, which are prevalent across biological and energy-related applications.

\end{document}